\documentstyle[12pt]{article}
\newcommand{\beqn}{\begin{eqnarray}}
\newcommand{\eeqn}{\end{eqnarray}}
\def\spose#1{\hbox to 0pt{#1\hss}}
\def\lsim{\mathrel{\spose{\lower 3pt\hbox{$\mathchar"218$}}
     \raise 2.0pt\hbox{$\mathchar"13C$}}}
\def\gsim{\mathrel{\spose{\lower 3pt\hbox{$\mathchar"218$}}
     \raise 2.0pt\hbox{$\mathchar"13E$}}}

\def\mpl{M_{\rm P}}

\def\d{\rm d}

\def\sinb{\sin \beta}
\def\cosb{\cos \beta}
\def\tanb{\tan \beta}
\def\msbar{\overline{MS}}
\def\barv{\overline{v}}
\def\barm{\overline{m}}

\def\vev#1{{\langle#1\rangle}}
\def\etal{ {\it et al.} }
\def\ie{ {\it i.e.} }
\def\eg{ {\it e.g.} }

\def\PRL#1#2#3{{\sl Phys. Rev. Lett.} {\bf #1}, #2 (#3)}
\def\PRD#1#2#3{{\sl Phys. Rev.} {\bf D#1}, #2 (#3)}
\def\PLB#1#2#3{{\sl Phys. Lett.} {\bf B#1}, #2 (#3)}
\def\PREP#1#2#3{{\sl Phys. Rep.} {\bf #1}, #2 (#3)}
\def\NPB#1#2#3{{\sl Nucl. Phys.} {\bf B#1}, #2 (#3)}

\def\gev{{\rm GeV }}

\def\ifmath#1{\relax\ifmmode #1\else $#1$\fi}

\begin{document}
\thispagestyle{empty}
\setcounter{page}{0}

\begin{flushright}
MPI-PhT/96-03\\

\today 
\end{flushright}
\bigskip
\bigskip
\bigskip
\bigskip
\bigskip

\begin{center}
{\Large\bf The Next-to-Minimal Coleman-Weinberg Model
}\\
\vspace{2em}
\large
Ralf Hempfling

{\it Max-Planck-Institut f\"ur Physik,
Werner-Heisenberg-Institut,}

{\it D--80805 M\"unchen, Germany}

\vspace{1.5ex}
{\it Email:} {\tt hempf@mppmu.mpg.de}
\end{center}
\bigskip
\bigskip
\bigskip

\begin{abstract}
\noindent
In the standard model (SM) the condition that the Higgs mass parameter
vanishes is stable under radiative corrections and yields a
theory that can be renormalized using dimensional regularization.
Thus, this model allows to predict the Higgs boson mass.
However, it is phenomenologically ruled out in its minimal version.
Here, we present a phenomenologically viable, minimal extension
which only includes an additional SM singlet and a U(1)$_X$ gauge symmetry.

\end{abstract}
\newpage

The particle content of the standard model of elementary
particle physics (SM) is minimal in that it only contains
particles that have already been observed plus 
one Higgs doublet needed to break the electroweak symmetry.
The Lagrangian describing the interactions of the theory
is obtained by forming all gauge-invariant and lorentz-invariant 
combinations of fields with dimensions less than four.
The coefficients are in general arbitrary factors
which have to be determined by experiment.

In order to obtain a renormalizable theory none
of these terms can be omitted as they are needed to cancel
divergent contributions coming from quantum correction.
The exception are terms whose absence enhances the symmetry of the
theory.
An example is chiral symmetry in the absence of a tree-level
mass term for one or more fermions\cite{cheng-li}.

The potential of the SM contains only one
parameter with dimension (mass)$^2$ [and none with dimension (mass)],
namely the Higgs mass parameter $\mu^2$.
From this mass term arises the 
most sever problem of the SM:
the hierarchy problem\cite{hierarchy}.
Since the condition $\mu^2 = 0$ is not protected by
any symmetry in the SM\footnote{Only in supersymmetric extensions of
the SM can scalar mass terms be absent\cite{susyreview}}
the large hierarchy
between the electroweak scale and the Planck scale
$\mu^2/\mpl \simeq 10^{-34}\ll 1$
can only be achieved by excessive fine-tuning.

Rather than to explain the smallness of $\mu^2$ it is 
may be conceptually more convincing to 
assume $\mu^2 = 0$ altogether.
The electroweak breaking in such a model can be achieved
by a negative Higgs self-coupling at some low scale $\Lambda$
due to the renormalization effects of the gauge couplings.
In this model, with one parameter less than the SM
(\ie\ $\mu^2 = 0$ or $\mu^2 \ll \Lambda^2$)
the Higgs mass (in units of the Higgs vacuum expectation value)
is determined by the gauge and Yukawa couplings.
This idea was first introduced by Coleman and Weinberg
in ref.~\cite{cw} were an upper limit on the Higgs mass
(the CW bound) of $m_h \lsim 10~\gev$
and implicitly an upper limit on the top quark mass
$m_t \lsim m_Z$ was established.
Unfortunately, both limits are by now
in contradiction with experiment\cite{refhiggs}\cite{reftop}.
Nontheless, the study of models with particular
conditions for the Higgs mass parameters
is of continued interest\cite{cut-off}\cite{cwnew}

In this letter, we will present a simple extension of the
CW model that is still phenomenologically viable.
We assume that the Higgs mass parameter is generated dynamically
as the vacuum expectation value (VEV) of a singlet field $S$.
The tree-level potential of our model without mass terms
can be written as
\beqn
V_0 = {\lambda_\phi\over 2}(\phi^\dagger \phi)^2
  + {\lambda_S\over 2}(S^\dagger S)^2
  -  \lambda_X (\phi^\dagger \phi)(S^\dagger S)\,,
\label{pot}
\eeqn
where $\phi$ denotes the SM Higgs doublet.
This potential has an additional U(1)$_X$ symmetry that transforms 
$S\rightarrow \exp(i\alpha)S$. By promoting this global symmetry
to a local symmetry we introduce a gauge coupling $g_X$
that can trigger spontaneous symmetry breaking.
In addition, we avoid the existence of
a massless goldstone boson.

The $\beta$ functions are
\beqn
16 \pi^2 \beta_\phi &=&  16 \pi^2 \beta_\phi^{SM} + \lambda_X^2\,,\nonumber\\
16 \pi^2 \beta_S &=&  5 \lambda_S^2 + 2\lambda_X^2+ {3\over 8}g_X^4
-32\pi^2 \lambda_S \gamma_S\,,\nonumber\\
16 \pi^2 \beta_X &=&  \lambda_X 
\left(3 \lambda_\phi + 2\lambda_S-2\lambda_X \right)
-16 \pi^2\lambda_X \left(\gamma_S+\gamma_\phi\right)\,,\nonumber\\
16 \pi^2 \beta_{g_X} &=&  {1\over 24}g_X^3\,.
\label{beta}
\eeqn
where the anomalous dimensions are
$\gamma_S = 3 g_X^2/64\pi^2$ and
$\gamma_\phi = (3 g^{\prime 2}+9 g^2 - 12 h_t^2)/64\pi^2$.

It is clear that the potential is flat in the
direction $\tan^2 \beta_0 \equiv \lambda_X/\lambda_\phi$
if
\beqn
\lambda_\phi \lambda_S = \lambda_X^2\,.
\label{critical}
\eeqn
In this case the VEV is determined by quantum corrections.
Let us assume that $\lambda_\phi \lambda_S - \lambda_X^2>0$
at the Planck scale $\mpl \simeq 10^{19}$.
At a scale $\Lambda<\mpl$ the effective coupling are obtained
by solving the renormalization group equations (RGEs) in eq.~\ref{beta}.
We find that $\lambda_\phi$ will converge for small $\Lambda$
to its infrared (IR) fixed point,
$\lambda_X$ [assumed to be small] will only change very slowly with $\Lambda$,
and $\lambda_S$ will continue to decrease for small $\Lambda$ and
sufficiently large $g_X$.
Thus, at some scale $\Lambda_c<\mpl$ eq.~\ref{critical}
will be satisfied.

\begin{figure}
\vspace*{13pt}
\vspace*{3.2truein}             
\includegraphics{fig1a.ps}
\includegraphics{fig1b.ps}
\caption{
contours of $\lambda_\phi(\Lambda)\lambda_S(\Lambda)=\lambda_X^2(\Lambda)$
(a) in the $\alpha_X$--$\ln \Lambda$ plane and
(b) in the $\lambda_S$--$\ln \Lambda$ plane}
\label{fig1}
\end{figure}

In fig.~\ref{fig1} we present contours for which
eq.~\ref{critical}
is satisfied (a) in the $\alpha_X(\mpl)$--ln$\Lambda$ plane
and (b) in the $\lambda_S(\mpl)$--ln$\Lambda$ plane.
We fix $\lambda_S(\mpl) = 0.1$,
$\lambda_\phi(\mpl) = 1$,
and $\alpha_X(\mpl) = 0.06$ whenever these
parameters are not varied.
Furthermore, we set $m_t = 176~\gev$ in all plots\cite{reftop}.
If there are two scales $\Lambda$ for which eq.~\ref{critical}
is satisfied then the larger (smaller)
one corresponds to a minimum (maximum).
This implies that there is an upper (lower) limit for $\lambda_S$
($\alpha_X$) as a function of $\alpha_X$
($\lambda_S$) for which the potential can have a minimum at the
electroweak scale.

Without any mass terms in the potential
it is convenient to transform to polar coordinates
$\phi \rightarrow \sinb r$ and $S \rightarrow \cosb r$
where $\tanb = \phi/S$.
At the scale $\Lambda_c$ the RG improved potential 
can be written as
\beqn
V_{RG} = r^4 \left[{\bar v}_0+{\beta_r\over 2} 
\ln\left({r^2\over \Lambda_c^2}\right)\right]\,,
\eeqn
with $\beta_r = \beta\phi \sin^4 \beta + \beta_S\cos^4 \beta 
- \beta_X\sin^2 2\beta/2$
and ${\bar v}_0 = \lambda_1\cos^4 \beta (\tan^2 \beta- \tan^2 \beta_0)^2/2$.
The Higgs mass matrix is given by
\beqn
{\cal M}_{i j}^2 = 
\left.{\d V_{RG} \over 2 \d v_i \d v_j}\right|_{r, \beta}\,,
\qquad v_1, v_2 = \phi, S\,,\label{mmatrix}
\eeqn
with $r$ and $\beta$
obtained by solving the minimum conditions
$\d V /\d r=0$ and
$\d V /\d \beta=0$.
The Higgs mass eigenvalues and mixing angles are 
\beqn
(m_{H/h}^2)_{RG} &=& {1\over 2}
\left({\rm tr}{\cal M}^2 
\pm \sqrt{{\rm tr}^2{\cal M}^2-4 {\rm det}\,{\cal M}^2}\right)\,,\nonumber\\
\sin 2\alpha_{RG} &=& {2 {\cal M}^2_{\phi S}\over {\rm tr}\,{\cal M}^2}
\eeqn

Unfortunately, the minimization of the potential yields rather
complicated expressions. The calculation can be simplified 
significantly by using the tree-level relation
$\alpha = \beta = \beta_0$.
This is justified by the observation
that any mixing effect of the radial degree of freedom $r$ 
with the remaining degree of freedom on the mass eigenvalues
will only be of second order and can be neglected.
This means that the lightest (heavies) mass eigenvalue $m_h$ ($m_H$)
is only determined by $V_{RG}$ ($V_0$).
Thus,  we can derive the approximate RG improved Higgs mass
\beqn
{\d \over \d r}V_{RG} &=& 2\beta_r
r^3\left[\ln\left({r^2\over \Lambda_c^2}\right)+{1\over 2}
+{\barv_0\over \beta_r} \right]=0\,,\label{dv}\\
(m_h^2)^{app}_{RG} &=& {\d^2 \over 2\d r^2}V_{RG} = 3\beta_r
r^2\left[\ln\left({r^2\over \Lambda_c^2}\right)+{7\over 6}
+{\barv_0\over \beta_r} \right]\nonumber\\
&=& 2 \beta_r r^2
 =  {3\alpha_X^2\over 8\sqrt{2}G_\mu}\cot^2 \beta 
\left[1+O(\lambda_X/\lambda_\phi)\right]\,.\label{rgapp}
\eeqn
Here, $\alpha_x = g_X^2/4\pi$ and
$G_\mu = (1.166\,39\pm0.000\,02)\times10^{-5}$ GeV$^{-2}$
is the fermi-constant\cite{pdg}.
Furthermore, we find that the Higgs vacuum expectation value
$r = \exp(-1/4)\Lambda_c$ is exponentially sensitive
to all the input parameters. However, a prediction of
$r$ in terms of $\lambda_i(\mpl)$ ($i = \phi, S, X$)
is not of interest for us but rather the value of the Higgs masses
in units of $r$.

\begin{figure}
\vspace*{13pt}
\vspace*{3.2truein}             
\includegraphics{fig2a.ps}
\includegraphics{fig2b.ps}
\caption{
comparison of (a) $m_h$ and (b) $\sin \alpha$ and $\sin \beta$
as a function of $\lambda_X$ using different approximations.}
\label{fig2}
\end{figure}

\begin{figure}
\vspace*{13pt}
\vspace*{6.6truein}             
\includegraphics{fig3a.ps}
\includegraphics{fig3c.ps}
\includegraphics{fig3b.ps}
\includegraphics{fig3d.ps}
\caption{Contours of constant (a) $m_h$, (b) constant $m_H$,
(c) constant $\cot \alpha$ and (d) constant $m_B$
in the $\alpha_X$--$\lambda_X$ plane.
In the shaded region the potential has maximum rather than a minimum
and the region above the dashed curve is ruled out by non-observation of
the process $Z\rightarrow h f\bar f$.}
\label{fig3}
\end{figure}

The renormalization group approach is very well suited to
evolve the effective theory over a large energy range 
and it provides in general very simple and transparent formulae.
Nonetheless, an independent check via a more complete
calculation is desirable.
Here, we will use the one-loop effective
potential
given in dimensional reduction\cite{dim-red} by\cite{effpot}
\beqn
V_1 = {r^4\over 64 \pi^2} \sum_\phi N_\phi \barm^4_\phi
\left(\ln{\barm^2_\phi}
-{3\over 2}-\Delta + \ln{r^2\over \Lambda_c^2}\right)\,.
\label{pot1}
\eeqn
Here, $N_\phi$ denotes the number of degrees of freedom
(with a $-$ sign for
fermions), $\barm_\phi$ stands for the masses of all the particles $\phi$ in
units of $r$
and $\Delta$ parameterizes the regularized divergences.
The one-loop divergences in eq.~\ref{pot1} are canceled by 
the divergences of the bare quantities.
In a modified minimal subtraction scheme ($\msbar$)
the renormalized quantities are obtained from the bare 
quantities by subtracting only the pieces proportional to $\Delta$.
Thus, we have to interpret the fields and couplings as bare quantities
(with superscript $0$) which are split into renormalized quantities  
(without superscript) plus counter-term, \ie\
\beqn
\lambda_i \rightarrow \lambda_i^0 &=& 
\lambda_i + \beta_{\lambda_i}\Delta\,,\nonumber\\
\Phi    \rightarrow \Phi^0    &=& 
\Phi\left(1 + {1\over 2}\gamma_\Phi\Delta \right)\,.\label{wavems}
\eeqn
Here, $i = \phi, X, S$ and $\Phi = \phi, S$.
Thus, the renormalized one-loop effective potential is obtained 
from eq.~\ref{pot} by interpreting all couplings and fields as 
$\msbar$ quantities and by setting $\Delta = 0$ in eq.~\ref{pot1}.
Note that we have removed all the one-loop divergences
in dimensional regularization\cite{dim-reg}
without requiring a mass counter-term.
The situation would be different had we used a cut-off
scheme [see e.g. ref.~\cite{cut-off}].
Clearly, if the cut-off is physical then a mass term of the order of the cut-off 
will be generated via radiative corrections in a non-supersymmetric model.
However, if the only scale is $\mpl$ and there is no new physics
[except gravity which presently can not be consistently combined with a
quantum field theory] then our results based on
dimensional regularization will
not be invalidated because it does not include a cut-off.

The minimization of the full one-loop potential is more
complicated and can only be done numerically.
The mass matrix is obtained from eq.~\ref{mmatrix}
by replacing $V_{RG}$ with $V_{1L} = V_0 + V_1$.
A convenient analytic result can again be obtained by 
neglecting mixing effects. Following the procedure of eq.~\ref{dv}--\ref{rgapp}
we obtain 
\beqn
(m_h^2)_{1L}^{app} = {r^2\over 16\pi^2}\sum_\phi \barm^4_\phi\,,
\label{1lapp}
\eeqn
We can easily check that the expressions for
$(m_h^2)_{RG}^{app}$ and  $(m_h^2)_{1L}^{app}$
are equivalent (they still differ numerically since in general
the place of the minimum is different, \ie: $\beta_{RG} \neq \beta_{1L}$).

The numerical comparison in fig.~\ref{fig2} reveils
an excellent agreement of the various methods 
under investigation here.
We present the prediction of $m_h$, $\sin\alpha$ and $\sin\beta$
using a RG and one-loop effective potential approach
and we compare $m_h$ obtained by diagonalizing the $2\times2$ mass matrix
with the second derivative in radial direction (denoted by subscript
$app$; of course in this approximation $\alpha \equiv \beta$).
Note that for values of $\lambda_X < 0.002$ the one-loop mass
corresponding to the field parallel to the VEV
becomes larger than the tree-level mass corresponding the the field
orthogonal to the VEV. 
Thus, for $\lambda_X\gsim 0.002$ the mass-eigenstate $h$ is predominantly the
CP-even, neutral component of the SM Higgs doublet
with $(m_h/r)^2\propto \lambda_X^{-1}$.
On the other hand, for $\lambda_X\lsim 0.002$
the mass-eigenstate $h$ is predominantly the
CP-even component of the SM Higgs singlet, $S$,
with constant mass $m_h$.
However, $(m_h)^{app}$ is defined as the
one-loop mass of the Higgs boson in radial direction and
will continue to rise with decreasing $\lambda_X$ even
if $\lambda_X < 0.002$

We will now determine the phenomenologically allowed
region in parameter space.
We assume that there is no new physics below $\mpl$ which
implies that the SM Higgs self coupling is very close to its IR fixed point
(we chose $\lambda_\phi = 1$).
The Higgs singlet self coupling is determined by fixing the scale 
of electroweak symmetry breaking
(\ie\ $\lambda_S = \lambda_X^2\lambda_\phi$).
We note that the singlet couples to the SM particles only
through mixing in the Higgs 
sector via $\lambda_X$.
This allows for a detection of the Higgs bosons via
$Z\rightarrow h f\bar f$ and $Z\rightarrow H f\bar f$.
The branching fractions of $h$ and $H$ into fermions and gauge bosons 
are the same as in the SM except for the possible decay $H\rightarrow hh$.
However, the decay widths $\Gamma(Z\rightarrow h f\bar f)$ and
$\Gamma(Z\rightarrow H f\bar f)$ are suppressed with respect to the SM
Higgs production by a factor $\sin^2 \alpha$ and  $\cos^2 \alpha$,
respectively.

In fig.~\ref{fig3} we present the particle spectrum of our model
in the $\alpha_X$--$\lambda_X$ plane in the RG approximation.
The phenomenologically interesting quantities are
the two CP-even Higgs masses [fig.~\ref{fig3} (a) and (b)],
the corresponding mixing angle $\cot \alpha$ [fig.~\ref{fig3} (c)]
and the mass of the U(1)$_X$ gauge boson $m_B$ [fig.~\ref{fig3} (d)].
The analysis of LEP data in ref.~\cite{sinba} for the two Higgs doublet
model is directly applicable to our case by replacing
$\sin^2 (\beta-\alpha)$ in favor of $\sin^2 \alpha$. 
The resulting limits are indicated in fig.~\ref{fig3}.
The area above the dashed curve is ruled out by non-discovery of 
a Higgs boson at LEP~\cite{sinba}.

To summarize, we have investigated a model for spontaneous
electro-weak symmetry breaking without mass parameters.
In the SM this scenario is ruled out by the large top quark mass
and we have demonstrated that it is still possible in models with
an extended Higgs sector. The model under investigation here
contains only an additional Higgs singlet and an additional U(1)$_X$
gauge symmetry needed to break the symmetry and to absorb
the related goldstone boson.
The main idea of the model is that the generation of the
mass term via the Coleman-Weinberg mechanism
occurs in a SM singlet sector and it is communicated to the
SM via mixing in the Higgs sector parameterized by $\lambda_X$.
The symmetry breaking is triggered by the U(1)$_X$ gauge symmetry.
Note that there is not Yukawa-type coupling of fermions to
the Higgs singlet $S$ that could spoil this mechanism.
The breaking of SU(2)$_L\times$U(1)$_Y$
gauge symmetry in our model is not triggered by the
Higgs self-coupling $\lambda_\phi$ turning negative at low scales
as is the case in the CW model, but in the conventional
fashion by a negative Higgs squared mass term $m^2_\phi = -\lambda_X\vev{S}^2$.
As a result, we find that the lightest Higgs boson mass is unconstrained
as long as this mixing with the SM Higgs boson is small
($\lambda_X \lsim 0.01$. The lower limit of the mass of the
U(1)$_X$ gauge boson is $m_B\gsim 250~\gev$.
The upper bound of the SM like Higgs boson is essentially 
the SM IR fixed point obtained in ref.~\cite{irfp}
possibly enhanced by as much as $20~\gev$ due to mixing effects
for large $\lambda_X$ and large $\alpha_X$.

\underline{Acknowledgement:} I would like to thank Dr. T. Hambye
for several useful discussions.


\end{document}